\documentclass[aps,pra,twocolumn,showpacs]{revtex4}

\usepackage{graphicx}
\usepackage{amsmath}
\usepackage{amssymb}

\begin{document}

\title{Reduced fidelity in Kitaev honeycomb model}
\author{Zhi Wang$^{1,2}$, Tianxing Ma$^{1,2,}$\footnote{txma@phy.cuhk.edu.hk}, Shi-Jian Gu${^1}$ and Hai-Qing
Lin${^1}$}
\address{$^{1}$Department of Physics and ITP,
The Chinese University of Hong Kong, Hong Kong \\
$^{2}$ Department of Physics, Beijing Normal University, Beijing
100875, China }

\begin{abstract}
We study reduced fidelity and reduced fidelity susceptibility in the
Kitaev honeycomb model. It is shown that the nearest-two-site
reduced fidelity susceptibility manifest itself as a peak at the
quantum phase transition point, although the one-site reduced
fidelity susceptibility vanishes. Our results directly reveal that
the reduced fidelity susceptibility can be used to characterize the
quantum phase transition in the Kitaev honeycomb model, which
suggests that, despite its local nature, the reduced fidelity
susceptibility is an accurate marker of the topological phase
transition when it is properly chosen.
\end{abstract}

\pacs{03.67.-a, 64.60.-i, 05.30.Pr, 75.10.Jm}
\date{\today}
\maketitle

\section{introduction}

The quantum phase transition (QPT), which is a phase transition
driven purely by quantum fluctuations and which occurs at zero
temperature,
is believed to be an important concept in condensed matter
physics\cite{continuous phase transition,quantum phase transition}.
It is defined as the nonanalytic behavior of the ground-state
properties, and thus reflects quantum fluctuations, which
differentiates it from the temperature-driven thermal phase
transition that reflects thermal fluctuations. Most QPTs can be
described within the traditional symmetry-breaking formalism;
however, there are also exceptions which can only be characterized
by topological order\cite{wen-book}. In these topological quantum
phase transitions (TQPTs), the local perturbation effects may be
exponentially suppressed, as was observed in fractional quantum Hall
phases\cite{wen-book}.

This QPT has appeared in a number of unrealistic models and is also
related to many realistic systems such as high temperature
superconductors\cite{PALee2006}. Despite many theoretical examples
that show the existence of the QPT, there is still no definite way
to mark it in terms of a local-order parameter. Recently, much
attention\cite{AHamma07,DFAbasto08,Yang2008,JHZhao0803,Erik2009,Buonsante1,TJ2009,PZanardi032109,PZanardi0701061,WLYou07,SChen07,SJGu072,MFYang07,NPaunkovic07}
has been drawn to the use of fidelity to mark the QPT ( for a
review, see Ref. \cite{Gurev} ).
For example, the fidelity approach to the TPQT occurring in the
Kitaev toric model have been used in Refs.
\cite{AHamma07,DFAbasto08}, and the fidelity between two states in
the Kitaev honeycomb model\cite{Kitaev2006} has been studied in
Refs. \cite{JHZhao0803,Yang2008}. E. Eriksson and H. Johannesson
proposed that several TQPTs were accurately signaled by a
singularity in the second-order derivative of the reduced
fidelity\cite{Erik2009}.
Fidelity is suggested to be particularly suited for revealing a QPT
in the Bose-Hubbard model\cite{Buonsante1}. Moreover, the
ground-state fidelity and various correlations to gauge the
competition between different orders in strong correlated system
were also discussed\cite{TJ2009}.

Fidelity is a concept borrowed from quantum-information theory and
is defined as a measure of similarity between two quantum states.
Intuitively, it would be a good marker, since the structure of the
ground state should suffer a dramatic change through the phase
transition point. As fidelity is purely a quantum-information
concept, the advantage of using fidelity is that no $a$ $priori$
knowledge of any order parameter and changes of symmetry of the
system need to be assumed. This idea has been used to analyze the
QPT, and its validity has been confirmed in many
systems\cite{Gurev}.

Meanwhile, the role of the leading term of fidelity was explored \cite%
{PZanardi0701061,WLYou07}. The fidelity susceptibility, which is the
second derivative of the fidelity with respect to the driving
parameter, was introduced to study the QPT\cite{WLYou07}. It has
been pointed out that the fidelity susceptibility is actually
equivalent to the structure factor (fluctuation) of the driving term
in the Hamiltonian on the perturbation level. In this case, if the
temperature is chosen as the driving parameter of thermal phase
transitions, the fidelity susceptibility, which is extracted from
the mixed-state fidelity between two thermal states\cite{WXG}, is
simply the specific heat. From this point of view, using the
fidelity to deal with the QPT seems still to be within the framework
of the correlation function approach, which is intrinsically related
to the local order parameter. Except for the global fidelity, which
measures the difference between ground states differing slightly in
the driving parameter, the reduced fidelity, which describes the
difference of the mix-state of only a local region of the system of
interest, was also introduced. It has been demonstrated that the
reduced fidelity should be useful in many ordinary symmetry-breaking
QPTs as well as some TQPTs\cite{Erik2009,WXG,Xiong2009}. The reduced
fidelity reveals information about a change in the inner structure
for a system undergoes a QPT, and it is significant to investigate
the behavior of the reduced fidelity susceptibility in both critical
and noncritical regions. Recently, it was shown that the TQPT in the
Kitaev spin model can be characterized by nonlocal-string order
parameters\cite{Feng2007,Chen2008,You2009}, so it would be
interesting to see whether it is able to mark this TQPT with a local
quantity; namely, the reduced fidelity.

In this work, we study the reduced fidelity and reduced fidelity
susceptibility in the Kitaev honeycomb model. Because of its
potential application in topological quantum computation, such a
model has been the focus of research in recent years, despite its
being rather
artificial\cite{Kitaev2006,Wen2003,RMP2008,Schmidt2008,You2009,Nussinov2009}.
We show that the reduced fidelity susceptibility of two nearest
sites manifest itself as a peak at the QPT point, although the
one-site reduced fidelity susceptibility vanishes. Our results
directly reveal that the reduced fidelity susceptibility can be used
to characterize the QPT in the Kitaev honeycomb model, and thus
suggest that the reduced fidelity susceptibility is an accurate
marker of the TQPT when it is properly chosen, despite its local
nature.

This article is organized as follows. The model and the calculation
of reduced fidelity are presented in Sec. II. The results of the
reduced fidelity and fidelity susceptibility are discussed in Sec.
III. A summary is given in Sec. IV.


\section{Reduced fidelity and fidelity susceptibility}

The Kitaev honeycomb model was first introduced by Kitaev to study
the topological order and anyonic statistics, which describe a
honeycomb lattice system  with a spin 1/2 located at each site. The
Hamiltonian consists of an anisotropic nearest-site interaction with
three types of bonds: $J_x$, $J_y$, and $J_z$. The model Hamiltonian
is
\begin{eqnarray}
H&=&-J_{x}\sum_{x\text{-bonds}}\sigma_{j}^{x}\sigma _{k}^{x}-J_{y}\sum_{y
\text{-bonds}}\sigma _{j}^{y}\sigma _{k}^{y}-J_{z}\sum_{z\text{-bonds}
}\sigma _{j}^{z}\sigma _{k}^{z}  \notag \\
&=&-J_xH_x-J_yH_y-J_zH_z,  \label{eq:Hamiltonian}
\end{eqnarray}
where $j$ and $k$ denote the two ends of the corresponding bond, and
$J_\alpha$ and $\sigma^\alpha(a=x,y,z)$ are dimensionless coupling
constants and Pauli matrices respectively.

Bellow, we solve this model within the method developed by
Kitaev\cite{Kitaev2006}. Here, we would like to mention that Z.
Nussinov and G. Ortiz have also illustrated how Kitaev's honeycomb
model may be solved directly\cite{Nussinov2009}. Within this
approach, the above Hamiltonian can be rewritten using fermionic
operators as
\begin{equation}
H=\sum_{\mathbf{q}}\sqrt{\epsilon _{\mathbf{q}}^{2}+\Delta _{\mathbf{q}}^{2}}%
\left( C_{\mathbf{q},1}^{\dag }C_{\mathbf{q},1}-C_{\mathbf{q},2}^{\dag }C_{%
\mathbf{q},2}\right) .
\end{equation}%
\bigskip with
\begin{eqnarray}
\epsilon _{\mathbf{q}} &=&J_{x}\cos q_{x}+J_{y}\cos q_{y}+J_{z},  \notag \\
\Delta _{\mathbf{q}} &=&J_{x}\sin q_{x}+J_{y}\sin q_{y}.
\end{eqnarray}
The momenta take the values
\begin{equation}
q_{x\left( y\right) }=\frac{2n\pi }{L},n=-\frac{L-1}{2},\cdots ,\frac{L-1}{2}
\end{equation}
where $L$ is  an odd integer; then the system size is $N=2L^{2}$.

Therefore, we have the ground state
\begin{equation}
\left\vert \Psi _{0}\right\rangle =\prod_{\mathbf{q}}C_{\mathbf{q},2}^{\dag
}\left\vert 0\right\rangle ,  \label{eq:ground}
\end{equation}%
with the ground-state energy
\begin{equation}
E_{0}=-\sum_{\mathbf{q}}\sqrt{\epsilon _{\mathbf{q}}^{2}+\Delta _{\mathbf{q}%
}^{2}}=-\sum_{\mathbf{q}}E_{\mathbf{q}}.  \label{eq:gse}
\end{equation}%
%
The two-site reduced density matrix can be expressed as \cite{You2009}
\begin{eqnarray}
\rho (\mathbf{r}_{1},\mathbf{r}_{2}) &=&\mathrm{Tr}^{\prime }\left(
|g\rangle \langle g|\right)  \notag \\
&=&\frac{1}{4}\sum_{\alpha ,\alpha ^{\prime }=0}^{3}\left\langle g\left\vert
\sigma _{{1}}^{\alpha }\sigma _{{2}}^{\alpha ^{\prime }}\right\vert
g\right\rangle \sigma _{{1}}^{\alpha }\sigma _{{2}}^{\alpha ^{\prime }},
\label{densitymatrix}
\end{eqnarray}%
where $\sigma ^{\alpha }$$($$\sigma ^{\alpha ^{\prime }}$$)$ are Pauli
matrices $\sigma ^{x}$, $\sigma ^{y}$, and $\sigma ^{z}$ for $\alpha $ $($$%
\alpha ^{\prime }$$)$$=1$ to 3, and the unit matrix for $\alpha $ $($$\alpha
^{\prime }$$)$$=0$.

Now we consider the nearest-two-site reduced density matrix, (i.e.,
the reduced density matrix of an $x$-bond. 
It can be proved that all but two parts of the density matrix are
nonzero:
\begin{eqnarray}
\rho (\mathbf{r},\mathbf{r+x})\!\!=\!\!\frac{1}{4}\left\langle g\left\vert
\sigma _{\mathbf{r}}^{x}\sigma _{\mathbf{r+x}}^{x}\right\vert g\right\rangle
\sigma _{\mathbf{r}}^{x}\sigma _{\mathbf{r+x}}^{x}\!+\frac{1}{4}I_{\mathbf{r}%
}I_{\mathbf{r+x}}.
\end{eqnarray}
If we assume the system has translational symmetry, we can calculate the
average
\begin{eqnarray}
\left\langle g\left\vert \sigma _{\mathbf{r}}^{x}\sigma _{\mathbf{r+x}%
}^{x}\right\vert g\right\rangle =\frac{1}{L^{2}}\left\langle g\left\vert
H_{x}\right\vert g\right\rangle =-\frac{1}{L^{2}}\frac{\partial E_{0}}{%
\partial J_{x}}.  \label{eq:rdm}
\end{eqnarray}
Combining Eqs. (\ref{eq:gse}) and (\ref{eq:rdm}), the reduced
density matrix can be expressed as
\begin{eqnarray}
C_{x} &\equiv &\left\langle g\left\vert \sigma _{\mathbf{r}}^{x}\sigma _{%
\mathbf{r+x}}^{x}\right\vert g\right\rangle  \notag \\
&=&-\frac{1}{L^{2}}\frac{\partial E_{0}}{\partial J_{x}}  \notag \\
&=&\frac{1}{L^{2}}\sum_{\mathbf{q}}\left( \frac{\epsilon _{\mathbf{q}}}{E_{%
\mathbf{q}}}\cdot \frac{\partial \epsilon _{\mathbf{q}}}{\partial J_{x}}+%
\frac{\Delta _{\mathbf{q}}}{E_{\mathbf{q}}}\cdot \frac{\partial \Delta _{%
\mathbf{q}}}{\partial J_{x}}\right)  \notag \\
&=&\frac{1}{L^{2}}\sum_{\mathbf{q}}\left( \frac{\epsilon _{\mathbf{q}}\cdot
\cos {q_{x}}+\Delta _{\mathbf{q}}\cdot \sin {q_{x}}}{E_{\mathbf{q}}}\right) .
\end{eqnarray}%
Similarly, for the $y$ and $z$ bonds, we can also obtain
\begin{eqnarray}
\rho(\mathbf{r},\mathbf{r+y})\! &\!=\!&\!\frac{1}{4}\left\langle g\left\vert
\sigma _{\mathbf{r}}^{y}\sigma _{\mathbf{r+y}}^{y}\right\vert g\right\rangle
\sigma _{\mathbf{r}}^{y}\sigma _{\mathbf{r+y}}^{y}\!+\frac{1}{4}I_{\mathbf{r}%
}I_{\mathbf{r+y}},  \notag \\
\rho (\mathbf{r},\mathbf{r+z})\! &\!=\!&\!\frac{1}{4}\left\langle
g\left\vert \sigma _{\mathbf{r}}^{z}\sigma _{\mathbf{r+z}}^{z}\right\vert
g\right\rangle \sigma _{\mathbf{r}}^{z}\sigma _{\mathbf{r+z}}^{z}\!+\frac{1}{%
4}I_{\mathbf{r}}I_{\mathbf{r+z}},  \label{xbond00}
\end{eqnarray}%
where
\begin{eqnarray}
C_{y}&\equiv&\left\langle g\left\vert \sigma _{\mathbf{r}}^{y}\sigma _{%
\mathbf{r+y}}^{y}\right\vert g\right\rangle  \notag \\
&=&-\frac{1}{L^{2}}\frac{\partial E_{0}}{\partial J_{y}}  \notag \\
&=&\frac{1}{L^{2}}\sum_{\mathbf{q}}\left( \frac{\epsilon _{\mathbf{q}}}{E_{%
\mathbf{q}}}\cdot \frac{\partial \epsilon _{\mathbf{q}}}{\partial J_{y}}+%
\frac{\Delta _{\mathbf{q}}}{E_{\mathbf{q}}}\cdot \frac{\partial \Delta _{%
\mathbf{q}}}{\partial J_{y}}\right)  \notag \\
&=&\frac{1}{L^{2}}\sum_{\mathbf{q}}\left( \frac{\epsilon _{\mathbf{q}}\cdot
\cos {q_{y}}+\Delta _{\mathbf{q}}\cdot \sin {q_{y}}}{E_{\mathbf{q}}}\right) ,
\notag \\
C_{z}&\equiv&\left\langle g\left\vert \sigma _{\mathbf{r}}^{z}\sigma _{%
\mathbf{r+z}}^{z}\right\vert g\right\rangle  \notag \\
&=&-\frac{1}{L^{2}}\frac{\partial E_{0}}{\partial J_{z}}  \notag \\
&=&\frac{1}{L^{2}}\sum_{\mathbf{q}}\left( \frac{\epsilon _{\mathbf{q}}}{E_{%
\mathbf{q}}}\cdot \frac{\partial \epsilon _{\mathbf{q}}}{\partial J_{z}}+%
\frac{\Delta _{\mathbf{q}}}{E_{\mathbf{q}}}\cdot \frac{\partial \Delta _{%
\mathbf{q}}}{\partial J_{z}}\right)  \notag \\
&=&\frac{1}{L^{2}}\sum_{\mathbf{q}}\left( \frac{\epsilon _{\mathbf{q}}}{E_{%
\mathbf{q}}}\right) .
\end{eqnarray}


After arriving at a diagonal reduced density matrix, it is easy and
straightforward to calculate the reduced fidelity
\begin{eqnarray}
F=\sum_i \sqrt{\rho_{ii} \rho^{\prime }_{ii}}.
\end{eqnarray}
Thus we have the expression
\begin{eqnarray}
F_\alpha&=&\frac{1}{2}[(1+C_\alpha)(1+C^{\prime
}_\alpha)+(1-C_\alpha)(1-C^{\prime }_\alpha)],
\end{eqnarray}
where $C_{\alpha}$ are $C_x$, $C_y$, and $C_z$ for $x$, $y$, and $z$
bond, respectively, and $F_{\alpha}$ are $F_x$, $F_y$, and $F_z$.

Now we calculate the reduced fidelity susceptibility $\chi_F$. For
diagonal density matrices, the fidelity susceptibility is
\begin{equation}
\chi_F=\sum_i\frac{(\partial_{J_{\alpha}} \rho_{ii})^2}{4\rho_{ii}}=\sum_i%
\frac{(\partial_{J_{s}} C_{\alpha})^2}{4\rho_{ii}},
\end{equation}
where $J_{s}$ is the driving parameter.

With this expression, we first set $J_{x}=J_{y}$ and select $J_{z}$
as the driving parameter, and then we have
\begin{eqnarray}
\epsilon _{\mathbf{q}}&=&\frac{1-J_{z}}{2}\cos q_{x}+\frac{1-J_{z}}{2}\cos
q_{y}+J_{z},  \notag \\
\epsilon _{\mathbf{q}}^{\prime }&\equiv &\frac{\partial \epsilon _{\mathbf{q}%
}}{\partial J_{z}}=-\frac{1}{2}\cos q_{x}-\frac{1}{2}\cos q_{y}+1,  \notag \\
\Delta _{\mathbf{q}}&=&\frac{1-J_{z}}{2}\sin q_{x}+\frac{1-J_{z}}{2}\sin
q_{y},  \notag \\
\Delta _{\mathbf{q}}^{\prime }&\equiv &\frac{\partial \Delta _{\mathbf{q}}}{%
\partial J_{z}}=-\frac{1}{2}\sin q_{x}-\frac{1}{2}\sin q_{y},  \notag \\
E_{\mathbf{q}}^{\prime }&\equiv &\frac{\partial E_{\mathbf{q}}}{\partial
J_{z}}=\frac{\epsilon _{\mathbf{q}}}{E_{\mathbf{q}}}\epsilon _{\mathbf{q}%
}^{\prime }+\frac{\Delta _{\mathbf{q}}}{E_{\mathbf{q}}}\Delta _{\mathbf{q}%
}^{\prime },
\end{eqnarray}%
and 
\begin{eqnarray}
\partial _{J_{z}}C_{x}&=&\frac{1}{L^{2}}\sum_{\mathbf{q}}\frac{1}{E_{\mathbf{%
q}}^{2}}  \notag \\
&&\lbrack (E_{\mathbf{q}}\epsilon _{\mathbf{q}}^{\prime }-E_{\mathbf{q}%
}^{\prime }\epsilon _{\mathbf{q}})\cos q_{x}+(E_{\mathbf{q}}\Delta _{\mathbf{%
q}}^{\prime }-E_{\mathbf{q}}^{\prime }\Delta _{\mathbf{q}})\sin q_{x}]
\notag \\
\partial _{J_{z}}C_{y}&=&\frac{1}{L^{2}}\sum_{\mathbf{q}}\frac{1}{E_{\mathbf{%
q}}^{2}}  \notag \\
&&\lbrack (E_{\mathbf{q}}\epsilon _{\mathbf{q}}^{\prime }-E_{\mathbf{q}%
}^{\prime }\epsilon _{\mathbf{q}})\cos q_{y}+(E_{\mathbf{q}}\Delta _{\mathbf{%
q}}^{\prime }-E_{\mathbf{q}}^{\prime }\Delta _{\mathbf{q}})\sin q_{y}]
\notag \\
\partial _{J_{z}}C_{z}&=&\frac{1}{L^{2}}\sum_{\mathbf{q}}\frac{E_{\mathbf{q}%
}\epsilon _{\mathbf{q}}^{\prime }-E_{\mathbf{q}}^{\prime }\epsilon _{\mathbf{%
q}}}{E_{\mathbf{q}}^{2}}.
\end{eqnarray}%

Also, we can set $J_z=1/3$ and make $J_x$ the driving parameter, and
the formula can be achieved in a similar way. Then, we can study the
reduced fidelity and reduced fidelity susceptibility depending on
which parameters we are interested in.


\section{Results and discussions}
\begin{figure}[tbp]
\includegraphics[width=7.5cm]{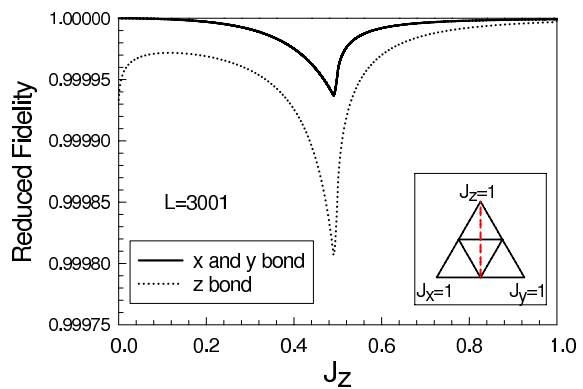}
\caption{(Color online) Two-site reduced fidelity as a function of
$J_z$ along the dashed line shown in the triangle for the $x$, $y$
and $z$ bonds.} \label{fig:rf1}
\end{figure}

\begin{figure}[tbp]
\includegraphics[width=7.5cm]{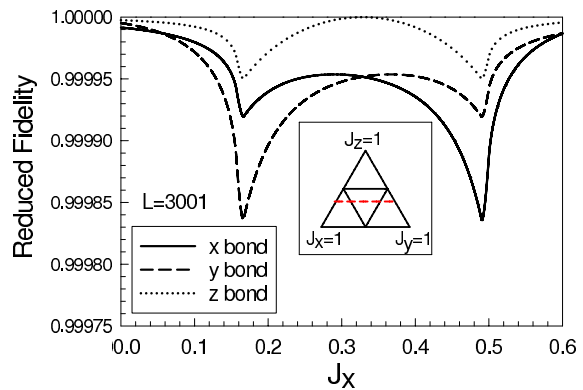}
\caption{(Color online) Two-site reduced fidelity as a function of
$J_x$ along the dashed line shown in the triangle for the $x$, $y$,
and $z$ bonds.} \label{fig:rf2}
\end{figure}

To study the reduced fidelity susceptibility of the Kitaev honeycomb
model, the parameters shall be restricted to the $J_x+J_y+J_z=1$
plane. According to Kitaev\cite{Kitaev2006}, this plane is divided
into a gapped phase and a gapless phase. These two phases are
separated by the triangle line connecting the $J_x=1/2$, $J_y=1/2$,
and $J_z=1/2$ points. As we already arrived at the expression of the
reduced fidelity and the reduced fidelity susceptibility, now we
begin to calculate these quantities numerically. We start from the
nearest-two-site reduced fidelity, since there are three types of
nearest sites for a honeycomb lattice. In Figs. \ref{fig:rf1} and
\ref{fig:rf2}, the two-site reduced fidelity with system size
$L=3001$ for all three different bonds are shown along two
orthogonal parameter directions, $J_x=J_y $ and $J_z=1/3$, which are
marked in the inset of the figure. In Fig. \ref{fig:rf1}, the
parameter is chosen to be $J_x=J_y$, and the reduced fidelities are
plotted as a function of $J_z$, where we can see that the $x$ bond
and $y$ bond reduced fidelities are the same, whereas the $z$-bond
reduced fidelity differs. In Fig. \ref{fig:rf2}, the parameter is
chosen to be $J_z=1/3$ and the reduced fidelities are plotted as
functions of $J_x$, where we can see that the $x-$ and $y-$bond
reduced fidelities are symmetric with respect to $J_x=1/3$, and the
$z-$bond reduced fidelity differs. However, in these two cases,
among all three bonds, the reduced fidelity manifests a dip at the
phase transition point. It is quite clear that the two-site reduced
fidelity can serve as a signature for the TQPT in the Kitaev
honeycomb model.

In order to illustrate the ability of reduced fidelity to mark the
TQPT in Kitaev honeycomb model, we calculate the reduced fidelity
susceptibility and present the results of the reduced fidelity
susceptibility for a system size $L$=3001 and for different bonds along $J_x=J_y$ and $J_z=1/3$ in Fig. \ref%
{fig:rfs1-xyz} and Fig. \ref{fig:rfs2-xyz}. As expected, the reduced
fidelity susceptibility shows a sharp peak at the TPQT point, which
would be a clear sign of the phase transition. Comparing with the
global fidelity susceptibility, the oscillation in the B phase seems
to be missing. Since it has been shown that the oscillating of the
global fidelity susceptibility might be related to the long-range
correlation function, this disappearance seems to represent the
locality of the bond-reduced fidelity susceptibility.
\begin{figure}[tbp]
\includegraphics[width=8cm]{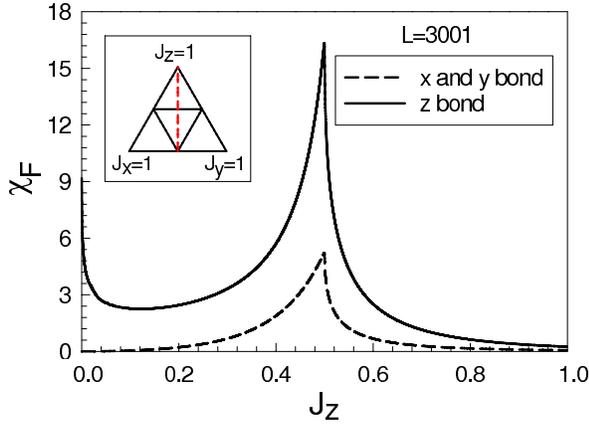}
\caption{(Color online) Two-site reduced fidelity susceptibility as
a function of $J_z$ along the dashed line shown in the triangle for
$x$, $y$, and $z$ bond.} \label{fig:rfs1-xyz}
\end{figure}

\begin{figure}[tbp]
\includegraphics[width=8cm]{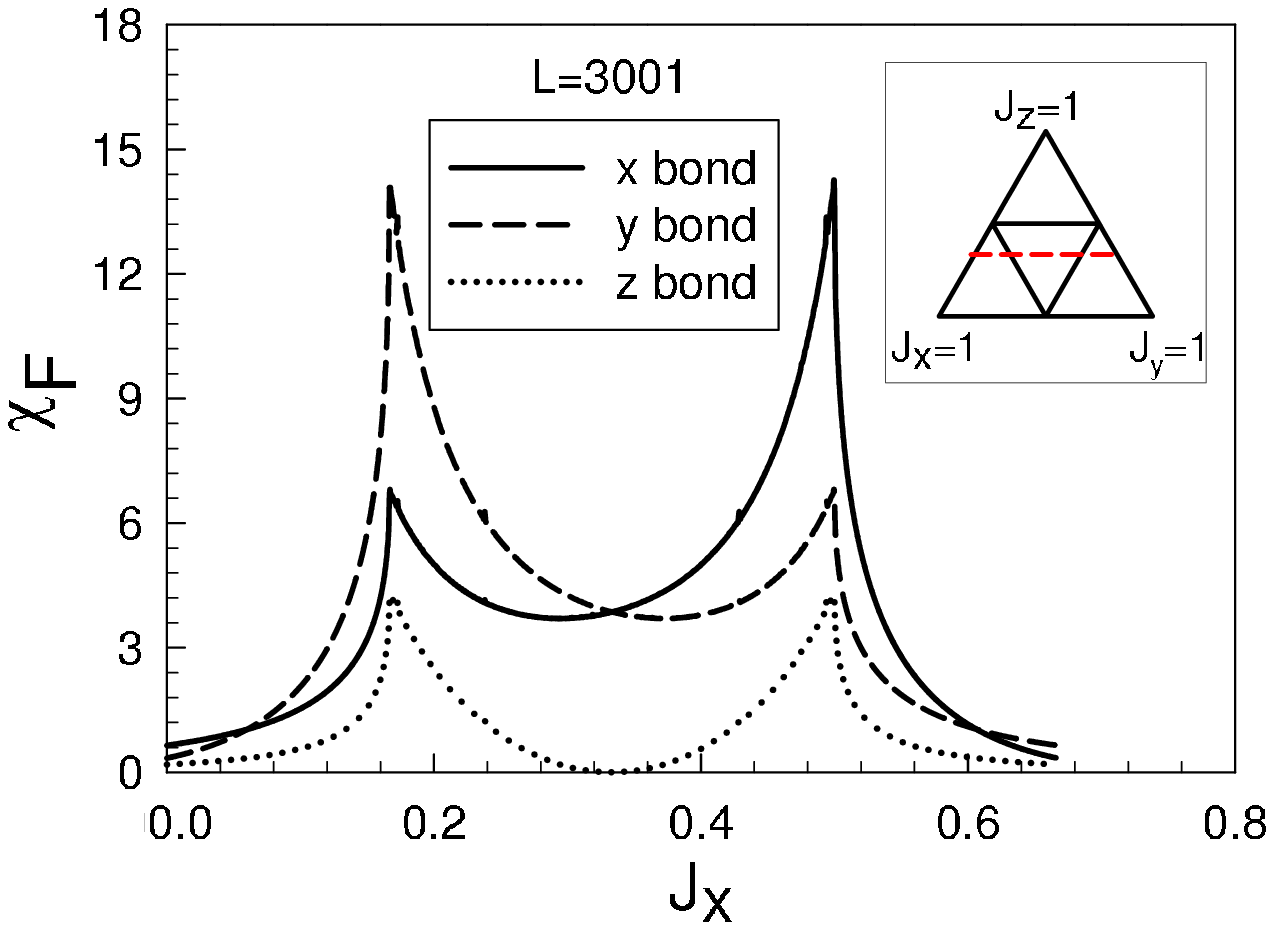}
\caption{(Color online) Two-site reduced fidelity susceptibility as
a function of $J_x$ along the dashed line shown in the triangle for
$x$, $y$, and $z$ bond.} \label{fig:rfs2-xyz}
\end{figure}

\begin{figure}[tbp]
\includegraphics[width=8cm]{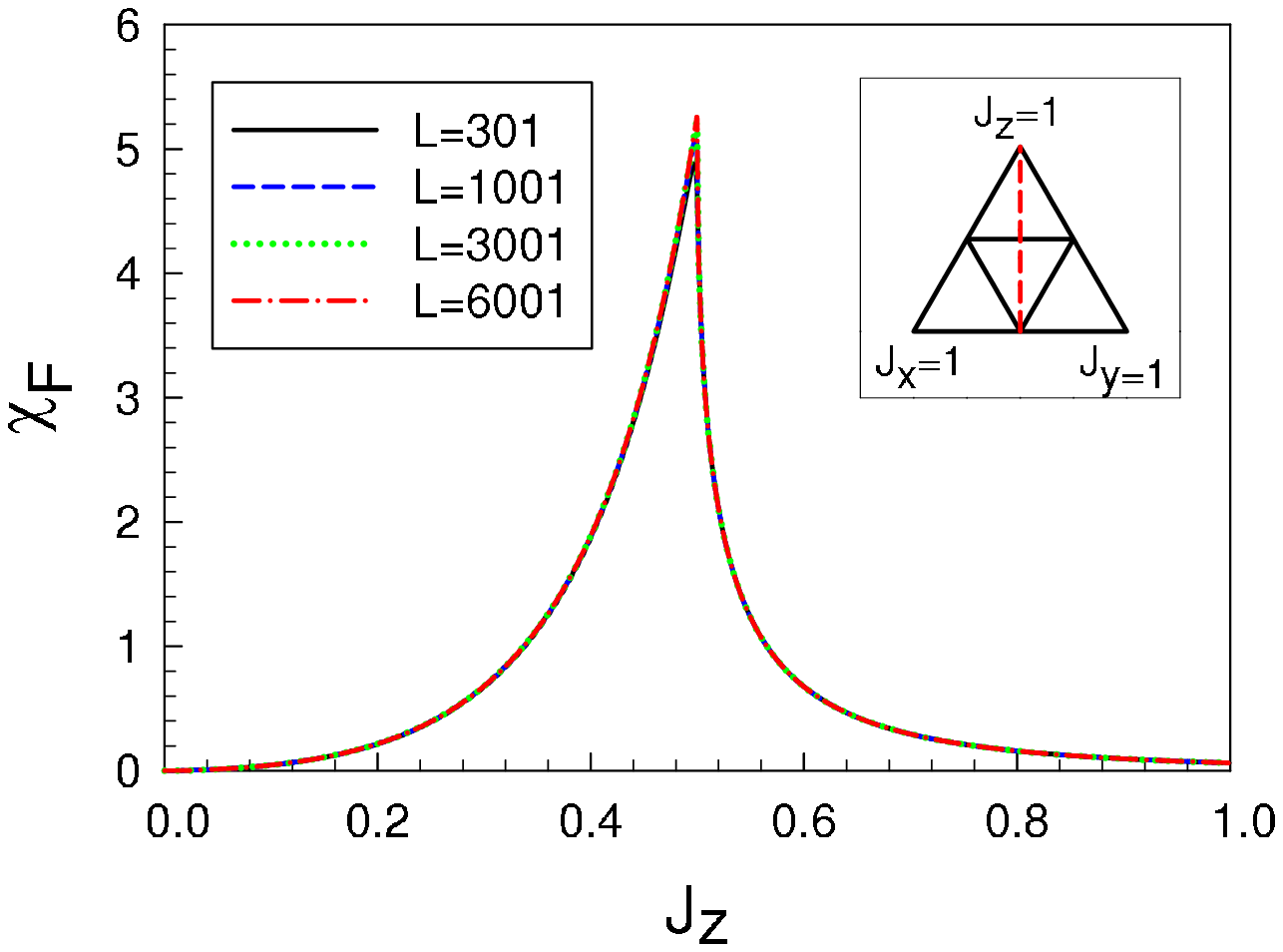}
\caption{(Color online) $x$-bond two-site reduced fidelity
susceptibility as a function of $J_z$ along the dashed line shown in
the triangle for the various system sizes $L$=301, 1001, 3001,
6001.} \label{fig:rfs1x-scaling}
\end{figure}

\begin{figure}[tbp]
\includegraphics[width=8cm]{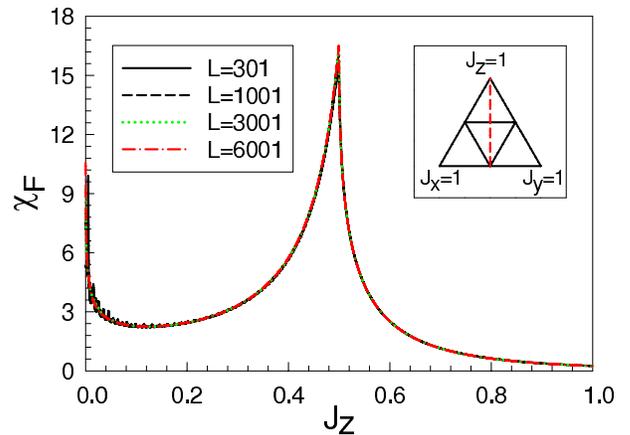}
\caption{(Color online) $z$-bond two-site reduced fidelity
susceptibility as a function of $J_z$ along the dashed line shown in
the triangle for the various system sizes $L$=301, 1001, 3001,
6001.} \label{fig:rfs1z-scaling}
\end{figure}

To further understand the properties of the reduced fidelity
susceptibility, we study its scaling behavior. In Figs. 5-9, the
reduced fidelity susceptibility of different bonds at different
parameter lines with different system sizes $L$ are
presented. Figure 5 ( Fig. 6 ) shows the $x-$bond ( $z-$bond ) two-site reduced fidelity susceptibility as a function of $%
J_z$, and Fig. 8 ( Fig. 9 ) shows the $x-$bond ( $z-$bond ) two-site
reduced fidelity susceptibility as a function of $J_x$, along the
dashed line shown in the triangle for the various system sizes
$L$=301, 1001, 3001, 6001. It is obvious that the susceptibility
barely changes with increasing lattice number in all the present
data.
\begin{figure}[tbp]
\includegraphics[width=8cm]{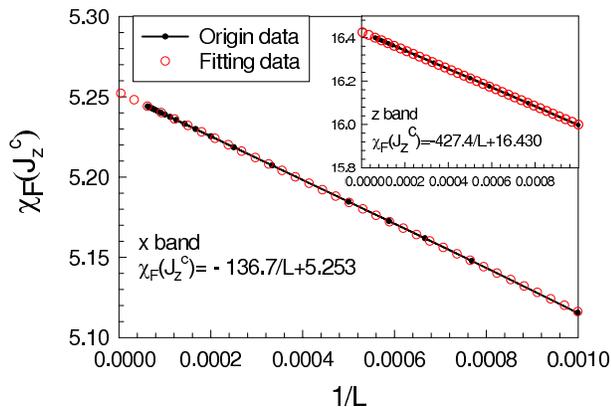}
\caption{(Color online) $x$-bond two-site reduced fidelity
susceptibility as a function of system sizes $L$ at the critical
point $J^{c}_z$=1/2; dark lines with points indicate numerical data
while open red circles indicate the fitting data. Inset: behavior of
$z$-bond two-site reduced fidelity
susceptibility as a function of the system size $L$ at the critical point $J^{c}_z$%
=1/2. }
\label{fig:rfs1x-scaling}
\end{figure}

From Figs. 5 and 6, one may also learn that the peak of reduced
fidelity increases slightly as the system sizes $L$ increases around
the critical point $J_{z}^{c}$=1/2. To study the scaling behavior of
the fidelity susceptibility at the critical point, we perform a
finite-size scaling analysis. In Fig. 7, dark lines with points
indicate $\chi _{F}(J_{z}^{c})(x)$, the $x-$bond two-site reduced
fidelity susceptibility at $J_{z}^{c}$ as a function of system size
$L$, and the red circles show the results of fitting data. In the
inset, the behavior of $\chi _{F}(J_{z}^{c})(z)$, the $z-$bond
two-site reduced fidelity susceptibility at $J_{z}^{c}$ is also
shown. The reduced fidelity susceptibility at the critical point
depends on the system size, which may be fit as
\begin{eqnarray}
\chi _{F}(J_{z}^{c})(x) &\simeq &-\frac{136.7}{L}+5.253,
\notag \\
\chi _{F}(J_{z}^{c})(z) &\simeq &-\frac{427.4}{L}+16.430.
\end{eqnarray}%
As shown in Fig. 7, the fitting data agree with the original data
rather well, so we may extrapolate $\chi _{F}(J_{z}^{c})$, in the
thermodynamic limit. In the thermodynamic limit scales, and for the
$x-$bond two-site reduced fidelity susceptibility, $\chi
_{F}(J_{z}^{c})$ is $5.253(\pm 0.001)$ , while $\chi
_{F}(J_{z}^{c})$ should be $16.430(\pm 0.001)$ for the $z-$bond
two-site reduced fidelity susceptibility.

\begin{figure}[tbp]
\includegraphics[width=8cm]{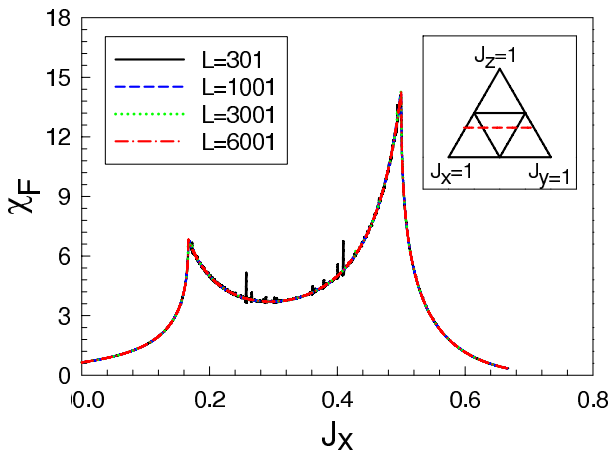}
\caption{(Color online) $x$-bond two-site reduced fidelity
susceptibility as a function of $J_x$ along the dashed line shown in
the triangle for the various system sizes $L$=301, 1001, 3001,
6001.} \label{fig:rfs2x-scaling}
\end{figure}

\begin{figure}[tbp]
\includegraphics[width=8cm]{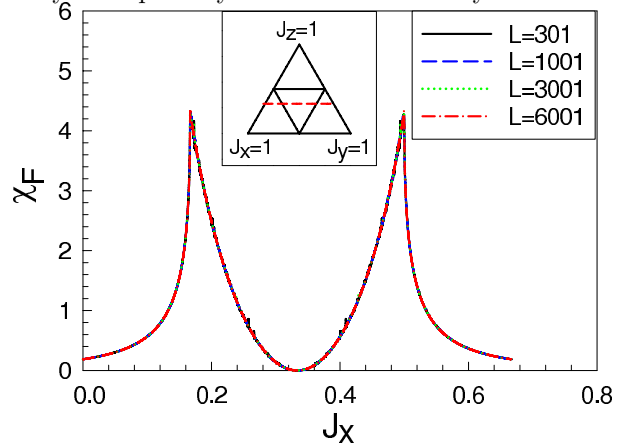}
\caption{(Color online) $z$-bond two-site reduced fidelity
susceptibility as a function of $J_x$ along the dashed line shown in
the triangle for various system sizes $L$=301, 1001, 3001, 6001. }
\label{fig:rfs2z-scaling}
\end{figure}

These results confirm the previous conjecture; namely, that, the
divergence of the global fidelity susceptibility is related to
long-range correlations. The divergence disappears in the reduce
fidelity susceptibility because the reduced fidelity susceptibility
averages over all long-range properties and only retains the
nearest-site correlation information.

\section{summary}

In summary, we study the reduced fidelity and reduced fidelity
susceptibility in the Kitaev honeycomb model. It is shown that the
 nearest-two-site reduced fidelity susceptibility manifest itself as a peak at the
quantum phase transition point, although the one-site reduced
fidelity susceptibility vanishes. Our results directly reveal that
the reduced fidelity susceptibility is able to mark the quantum
phase transition in the Kitaev honeycomb model, and thus suggest
that the reduced fidelity susceptibility is still an accurate marker
of the TPQT when it is properly chosen, despite its local nature.
The conclusion that such a local quantity can characterize a TQPT is
conceptually consistent with the fact that any physical observable
is local in nature.

\begin{acknowledgments}
The authors thank Wing-Chi Yu for helpful discussions. This work is
supported by the Earmarked Grant Research from the Research Grants
Council of HKSAR, China (Project No. HKUST3/CRF/09).
\end{acknowledgments}

\end{document}